\documentclass[10pt]{article}
\usepackage{amssymb,amsmath,mathrsfs}
\usepackage{graphicx,rotate,subfigure}
\usepackage{cite}
\usepackage[colorlinks=true,
            linkcolor=red,
            urlcolor=blue,
            citecolor=blue]{hyperref}

\usepackage{color}
\long\def\rpl#1!!#2!!{\textcolor{red}{#1} \textcolor{blue}{#2}}
\newcommand{\specialcell}[2][c]{%
  \begin{tabular}[#1]{@{}c@{}}#2\end{tabular}}

\def \order(#1){{\cal O} \left(#1 \right)}

\evensidemargin=\oddsidemargin
\def\Eqn#1{Eq.\ (\ref{#1})}

\begin{document}

\begin{flushright}
SINP/TNP/2014/07
\end{flushright}

\begin{center}
{\Large \bf Nondecoupling of charged scalars in Higgs decay to two
  photons \\and symmetries of the scalar potential} \\
\vspace*{1cm} {\sf Gautam
  Bhattacharyya\footnote{gautam.bhattacharyya@saha.ac.in}, ~ Dipankar
  Das\footnote{d.das@saha.ac.in}} \\
\vspace{10pt} {\small } {\em Saha Institute of Nuclear
    Physics, 1/AF Bidhan Nagar, Kolkata 700064, India}

\normalsize
\end{center}

\begin{abstract}
A large class of two- and three-Higgs-doublet models with discrete
symmetries has been employed in the literature to address various
aspects of flavor physics. We analyse how the precision measurement of
the Higgs to diphoton signal strength would severely constrain these
scenarios due to the nondecoupling behavior of the charged scalars, to
the extent that in the presence of exact discrete symmetry, the number
of additional non-inert scalar doublets can be constrained no matter
how heavy the nonstandard scalars are. We demonstrate that if the
scalar potential is endowed with appropriate global continuous
symmetries together with soft breaking parameters, decoupling can be
achieved thanks to the unitarity constraints on the mass-square
differences of the heavy scalars.

\end{abstract}

\bigskip

\section{Introduction}
The behavior of the scalar boson observed at the CERN Large Hadron
Collider (LHC) is tantalizingly close to that of the Standard Model
(SM) Higgs boson. A very timely and relevant question is whether this
scalar is the only one of its type as predicted by the SM, or it is
the first to have been discovered in a family of more such species
arising from an underlying extended scalar sector.  A natural
extension of the SM is realized by adding more SU(2) scalar doublets,
which we consider in this paper. There are two advantages for choosing
doublets.  First, the $\rho$-parameter remains unity at tree
level. Second, it is straightforward to find a combination, namely,
\begin{eqnarray}
h=\frac{1}{v}\sum\limits_{i=1}^{n} v_ih_i \,, ~~~~{\rm with}~~
v^2=\sum\limits_{i=1}^{n}v_i^2=(246~{\rm GeV})^2\,,
\label{SMh}
\end{eqnarray}
($v_i$ is the vacuum expectation value (vev) of the $i$-th doublet and
$h_i$ is the corresponding real scalar field), which has SM-like
couplings with fermions and gauge bosons. This is not in general a
mass eigenstate.  But when we demand that this is indeed the physical
state observed at the LHC with a mass $m_h \approx$ 125 GeV, we are
automatically led to the so called {\em alignment limit}. This limit
is motivated by the LHC data on the Higgs boson signal strengths in
different channels which are showing increasing affinity towards the
SM predictions.  In this paper we pay specific attention to the $h \to
\gamma\gamma$ process. Though this process is loop driven and has a
small branching ratio, it played an important r\^ole in the Higgs
discovery. Importantly, this branching ratio is expected to be
measured in LHC-14 with much greater accuracy.  Now, additional SU(2)
scalar doublets would bring in additional states, both charged and
neutral, in the spectrum. Here our primary concern is how those
charged scalars couple to $h$ and how much they contribute to the $h
\to \gamma \gamma$ rate as virtual states in loops. This leads to the
observation that even when the masses of the charged scalars floating
in the loop are taken to very large values, they do not {\em
  necessarily} decouple from this process. Deciphering the underlying
reasons behind this constitutes the motive of this paper. Although
this has been noted in the past in the context of two-Higgs-doublet
models (2HDM), only some cursory remarks were made on it without
exploring its full implications~\cite{Djouadi:1996yq, Arhrib:2003ph,
  Chang:2012ta, Bhattacharyya:2013rya,
  Ferreira:2014naa,Fontes:2014tga}.  We investigate the r\^ole of
symmetries which are imposed on the scalar potential in figuring out
under what conditions the decoupling of heavy charged scalars in the
$h \to \gamma \gamma$ loop takes place. The upshot is that if the
potential has an exact $Z_2$ symmetry {\em and} both the scalars
receive vevs, which is the case for a large class of 2HDM
scenarios~\cite{Branco:2011iw}, the contribution of the charged scalar
does not decouple. If $Z_2$ is softly broken by a term in the
potential then decoupling can be achieved at the expense of tuning of
parameters.  On the other hand, a global U(1) symmetry followed by its
soft breaking can ensure decoupling.  For simplicity, we first
demonstrate this behavior in the context of 2HDM.  We then address the
same question, for the first time, in the context of
three-Higgs-doublet models (3HDM). It is not difficult to foresee what
happens if we add more doublets, which leads us to draw an important
conclusion: unless decoupling is ensured, e.g. as we did by imposing a
global U(1) symmetry in the 2HDM potential, precision measurements of
$h \to \gamma \gamma$ branching ratio can put constraints on the
number of additional non-inert scalar doublets regardless of how heavy
the charged scalars are.  We recall that only lower bounds on charged
scalar masses have been placed from processes like $b \to s \gamma$,
as the effects decouple when their masses are heavy for all such
flavor observables.  Thus, precision measurements of $h \to
\gamma\gamma$ would provide complementary information.  Incidentally,
whatever we comment on $h \to \gamma \gamma$ applies for $h \to Z
\gamma$ as well at least on a qualitative level.

It should be noted that in multi-doublet scalar models, the production
cross section as well as the tree-level decay widths of the Higgs
boson remain unaltered from their respective SM expectations in the
alignment limit. Only the loop induced decay modes like
$h\to\gamma\gamma$ and $h\to Z\gamma$ will pick up additional
contributions induced by virtual charged scalars. However, the
branching ratios into these channels are too tiny compared to other
dominant modes. As a result, the total Higgs decay width will be
hardly modified. Considering all these, the expression for the
diphoton signal strength is simplified to
\begin{eqnarray}
 \mu_{\gamma\gamma} &\equiv & {\sigma(pp\to h) \over \sigma^{\rm SM}(pp\to
   h)} \cdot {\mbox{BR} (h \to \gamma\gamma) \over \mbox{BR}^{\rm SM}
   (h \to \gamma\gamma)}~=~\frac{\Gamma(h \to
   \gamma\gamma)}{\Gamma^{\rm SM}(h\to \gamma\gamma)} \, .
\label{mugg}
\end{eqnarray}
For convenience, we parametrize the coupling of $h$ to the charged
scalars as
\begin{eqnarray}
g_{hH_{i}^{+}H_{i}^{-}}=\kappa_{i}\frac{gm_{i+}^{2}}{M_{W}} \,,
\label{defkappa}
\end{eqnarray}
where $m_{i+}$ is the mass of the $i$-th charged scalar
($H_i^\pm$). As we will see later, the decoupling or nondecoupling
behavior of the $i$-th charged scalar from $\mu_{\gamma\gamma}$ is
encoded in $\kappa_i$.  The expression of the diphoton decay width of
the Higgs is given by \cite{Gunion:1989we}~:
\begin{eqnarray}
 \Gamma (h\to \gamma\gamma) = \frac{\alpha^2g^2}{2^{10}\pi^3}
 \frac{m_h^3}{M_W^2} \Big|\mathcal{A}_W + \frac{4}{3}\mathcal{A}_t +
 \sum_{i}\kappa_{i} \mathcal{A}_{i+} \Big|^2 \,,
\label{h2gg}
\end{eqnarray}
where, using $\tau_x \equiv (2m_x/m_h)^2$,
the expressions for $\mathcal{A}_W$, $\mathcal{A}_t$ and
$\mathcal{A}_{i+}$ are given by
%
\begin{eqnarray}
 \mathcal{A}_W = 2+3\tau_W+3\tau_W(2-\tau_W)f(\tau_W) \,, ~
 \mathcal{A}_t = -2\tau_t \big[1+(1-\tau_t)f(\tau_t)\big] \,, ~
 \mathcal{A}_{i+} = -\tau_{i+} \big[ 1-\tau_{i+}f(\tau_{i+}) \big] \,.
\end{eqnarray}
%
Since we are concerned with heavy charged scalars, we can take
$\tau_{x} > 1$ for $x = (W,~t,~H_{i}^{\pm})$, and then
$f(\tau) =
\left[\sin^{-1}\left(\sqrt{1/\tau}\right)\right]^2$.
Now plugging Eq.~(\ref{h2gg}) into Eq.~(\ref{mugg}), we obtain
\begin{eqnarray}
 \mu_{\gamma\gamma} &=& \frac{\Big|\mathcal{A}_W + \frac{4}{3}\mathcal{A}_t  +
    \sum_{i}\kappa_{i} \mathcal{A}_{i+}\Big|^2}{\Big|\mathcal{A}_W +
   \frac{4}{3}\mathcal{A}_t \Big|^2} \, .
\end{eqnarray}
In the limit the charged scalar is very heavy, the quantity
$\mathcal{A}_{i+}$ saturates to $1/3$.  If $\kappa_i$ also saturates
to some finite value in that limit then the charged scalar would not
decouple from the $h \to \gamma\gamma$ loop. Then no matter how heavy
the charged scalar is, $\mu_{\gamma\gamma}$ will differ from its SM
value.  If the experimental value of $\mu_{\gamma\gamma}$ eventually
settles on very close to the SM prediction then such nondecoupling
scenarios will be disfavored. The decoupling would happen only if
$\kappa_i$ falls with increasing charged scalar mass.  In what
follows, we will illustrate these features by considering some popular
doublet extensions of the SM scalar sector.

\section{Two Higgs-doublet models} 
We consider a 2HDM with $\phi_1$ and $\phi_2$ as the two scalar
doublets. Then we impose a $Z_2$ symmetry in the potential, namely,
$\phi_1 \to \phi_1$ and $\phi_2 \to - \phi_2$, to avoid Higgs mediated
flavor-changing neutral current in the fermionic sector.  The
expression of the scalar potential is displayed below~\cite{Gunion:1989we}~:
\begin{eqnarray}
 V_{\rm 2HDM} &=& 
 \lambda_1 \left( \phi_1^\dagger\phi_1 - \frac{v_1^2}{2} \right)^2 
+\lambda_2 \left( \phi_2^\dagger\phi_2 - \frac{v_2^2}{2} \right)^2 
+\lambda_3 \left( \phi_1^\dagger\phi_1 + \phi_2^{\dagger}\phi_2 
- \frac{v_1^2+v_2^2}{2} \right)^2
\nonumber \\*
&&  
+ \lambda_4 \left(
(\phi_1^{\dagger}\phi_1) (\phi_2^{\dagger}\phi_2) -
(\phi_1^{\dagger}\phi_2) (\phi_2^{\dagger}\phi_1)
\right) 
+ \lambda_5 \left( {\rm Re} ~ \phi_1^\dagger\phi_2 - \frac{v_1 v_2}{2} \right)^2 
+ \lambda_6 \left( {\rm Im} ~ \phi_1^\dagger\phi_2 \right)^2 \,,
\label{2hdm potential}
\end{eqnarray}
where the $\lambda_5$ term arises due to soft breaking of $Z_2$.  We
assume all the lambdas to be real, i.e., CP is not broken
explicitly. It is also implicitly assumed that both the scalar
doublets receive vevs.

First, it is important to count the number of free parameters. As we
have assumed the parameters to be real, there are only 8 free
parameters. Two of them $v_1$ and $v_2$ can be traded for $v$ and
$\tan\beta \equiv v_2/v_1$. All the remaining parameters, except
$\lambda_5$, can be traded for four physical scalar masses
($m_h,~m_H,~m_A,~m_{1+} (\equiv m_{H^+})$) and the rotation angle
($\alpha$) in the neutral CP even sector. The lambdas can be expressed
in terms of physical masses as:
\begin{subequations}
\begin{eqnarray}
\lambda_1 &=& \frac{1}{2v^2\cos^2\beta}\left[m_H^2\cos^2\alpha
  +m_h^2\sin^2\alpha
  -\frac{\sin\alpha\cos\alpha}{\tan\beta}\left(m_H^2-m_h^2\right)\right]
-\frac{\lambda_5}{4}\left(\tan^2\beta-1\right) \,, \\
\lambda_2 &=& \frac{1}{2v^2\sin^2\beta}\left[m_h^2\cos^2\alpha
  +m_H^2\sin^2\alpha
  -\sin\alpha\cos\alpha\tan\beta\left(m_H^2-m_h^2\right) \right]
-\frac{\lambda_5}{4}\left(\cot^2\beta-1\right) \,, \\
\lambda_3 &=& \frac{1}{2v^2}
\frac{\sin\alpha\cos\alpha}{\sin\beta\cos\beta}
\left(m_H^2-m_h^2\right) -\frac{\lambda_5}{4} \,, \\
\lambda_4 &=& \frac{2}{v^2} m_{1+}^2 \,, \\
\lambda_6 &=& \frac{2}{v^2} m_A^2 \,.
\end{eqnarray}
\label{inv2HDM}
\end{subequations}
The quantities that appear on the right-hand-side of \Eqn{inv2HDM} are
all {\em independent} parameters.  Since we work in the alignment
limit, it follows that $\alpha~(=\beta-\pi/2)$.  This means that we deal
with 7 independent parameters, out of which 5 are unknown, namely,
$~m_H,~m_A,~m_{1+},~\tan\beta$ and $\lambda_5$, while 2 are known, namely, $v
= 246$ GeV and $m_h \approx 125$ GeV.  As mentioned earlier, the
SM-like Higgs boson is recovered in this limit as shown in \Eqn{SMh}.

\begin{figure}
\begin{center}
\includegraphics[scale=0.35]{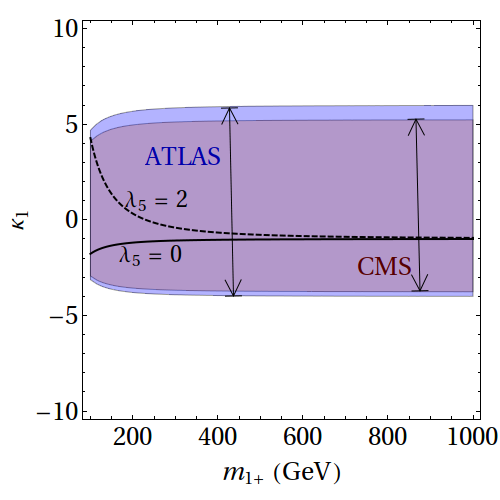} ~~~~~
\includegraphics[scale=0.35]{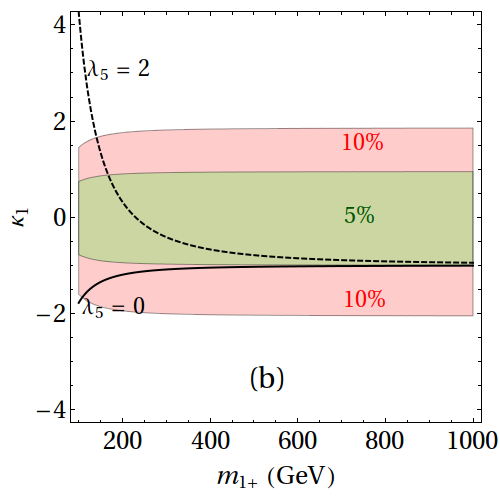}
\end{center} 
\caption{\em In the left panel (a) we display the constraints on
  $\kappa_1$ in 2HDM coming from the measured values of
  $\mu_{\gamma\gamma}$ at 95\% C.L. by the CMS ($1.14
  ^{+0.26}_{-0.23}$ \cite{Khachatryan:2014ira}) and ATLAS ($1.17
  \pm 0.27$ \cite{Aad:2014eha}) Collaborations. In the right panel
  (b) we show what would be the 95\% C.L. allowed range of $\kappa_1$
  if $\mu_{\gamma\gamma}$ is hypothetically measured to be as $1 \pm
  0.1 (0.05)$ in future colliders.  In both panels we have plotted
  Eq.~(\ref{kappaz2}) for two different values of $\lambda_5$.}
\label{fig1}
\end{figure}

The charged scalar contribution to $\mu_{\gamma\gamma}$ is controlled
by (putting $i=1$ in $\kappa_i$)~\cite{Djouadi:1996yq, Arhrib:2003ph,
  Bhattacharyya:2013rya, Ferreira:2014naa, Swiezewska:2012eh}
\begin{eqnarray}
\kappa_1 =
-\frac{1}{m_{1+}^2}\left(m_{1+}^2+\frac{m_h^2}{2}-\frac{\lambda_5
  v^2}{2} \right) \, .
\label{kappaz2}
\end{eqnarray}
Clearly, $\kappa_1$ saturates to $-1$ as the charged scalar becomes
excessively heavy. Decoupling can be achieved by tuning $m_{1+}^2
\simeq \lambda_5 v^2/2$ \cite{Gunion:2002zf}. Recalling our counting
of independent parameters, any adjustment between the charged scalar
mass and $\lambda_5$ is nothing short of fine-tuning.  On the other
hand, if the $Z_2$ symmetry in the scalar potential is exact, {\it
  i.e.}  $\lambda_5=0$, then the charged scalar will never decouple
and will cause $\mu_{\gamma\gamma}$ to settle below its SM prediction.
In Fig.~\ref{fig1} we have plotted the allowed range of $\kappa_1$ in
2HDM from the present LHC data as well as from an anticipation of
future sensitivity.

An interesting possibility arises when we employ a U(1) symmetry,
rather than the usual $Z_2$ symmetry, in the potential. The choice
$\lambda_5=\lambda_6$ will ensure U(1) symmetry in the quartic
terms. The bilinear term involving $\lambda_5$ still breaks the U(1)
symmetry softly.  Then the mass of the pseudoscalar gets related to
the soft breaking parameter $\lambda_5$ as $m_A^2 = \lambda_5 v^2/2 $.
In this case, the expression for $\kappa_1$ reads \cite{Bhattacharyya:2013rya}~:
\begin{eqnarray}
\kappa_1 = -\frac{1}{m_{1+}^2}\left(m_{1+}^2-m_A^2+\frac{m_h^2}{2} \right) \,.
\label{u1kappa}
\end{eqnarray}
In a previous paper~\cite{Bhattacharyya:2013rya}, we provided a
detailed analysis on the unitarity and stability constraints on
various combination of $\lambda_i$ couplings when the 2HDM scalar
potential has a softly broken U(1) symmetry. We cite some of them here
to demonstrate `decoupling' for large individual quartic couplings, as
what is constrained from unitarity is only their differences in
certain combinations. For example, $(2 \lambda_3 + \lambda_4) \leq
16\pi$ implies $(2 m_{1+}^2 - m_H^2 - m_A^2 +m_h^2) \leq 16\pi
v^2$. Also, $(\lambda_1 + \lambda_2 + 2 \lambda_3) \leq 16\pi/3$
implies $(m_H^2 - m_A^2)(\tan^2\beta + \cot^2\beta) + 2m_h^2 \leq
32\pi v^2/3$. These relations, together with $|m_{1+} - m_H| \ll
(m_{1+}, m_H)$ arising from the electroweak $T$ parameter, restrict
the splitting between the charged scalar and the pseudoscalar mass
($|m_{1+}^2-m_A^2|$). As displayed through more such relations among
quartic couplings and the associated plots in the plane of non-SM
scalar masses in \cite{Bhattacharyya:2013rya}, the individual scalar
masses can become very large without violating unitarity as long as
their mass-square differences are within certain limits.
Consequently, the numerator in \Eqn{u1kappa} cannot grow indefinitely
with increasing $m_{1+}$.  Thus $\kappa_1$ becomes very small in that
limit and $\mu_{\gamma\gamma}$ reaches the SM predicted value.  The
key issue is that the $Z_2$ symmetry breaking $\lambda_5$ term was not
related to the mass of any particle in the spectrum, and hence its
adjustment {\em vis-\`a-vis} the charged scalar mass was nothing short
of fine-tuning.  Now, the global U(1) breaking $\lambda_5$ is related
to the pseudoscalar mass whose splitting with the charged scalar mass
is restricted from unitarity.

\subsection{Underlying dynamics behind decoupling}
We now discuss the underlying reason behind decoupling or
nondecoupling of nonstandard scalars from physical processes in the
2HDM context. The conclusion is equally applicable for $n$HDM where $n
> 2$.  First, we write down the 2HDM potential using a different
notation:
\begin{eqnarray}
V_{\rm 2HDM} &=& m_{11}^2 \phi_1^\dagger\phi_1 +
m_{22}^2\phi_2^\dagger\phi_2 -\left(m_{12}^2 \phi_1^\dagger\phi_2
+{\rm h.c.} \right) +\frac{\beta_1}{2} \left(\phi_1^\dagger\phi_1
\right)^2 +\frac{\beta_2}{2} \left(\phi_2^\dagger\phi_2 \right)^2
\nonumber \\ && +\beta_3 \left(\phi_1^\dagger\phi_1 \right)
\left(\phi_2^\dagger\phi_2 \right) +\beta_4 \left(\phi_1^\dagger\phi_2
\right) \left(\phi_2^\dagger\phi_1 \right) +\left\{\frac{\beta_5}{2}
\left(\phi_1^\dagger\phi_2 \right)^2 +{\rm h.c.} \right\} \, .
\label{notation1}
\end{eqnarray} 
This parametrization does not {\em a priori} assume, unlike the one
used in this paper given in Eq.~(\ref{2hdm potential}), that $\phi_1$
or $\phi_2$ necessarily acquires any vev. In this parametrization, in
the limit when the dimensionless couplings $\beta_2 = \beta_3 =
\beta_4 = \beta_5 = 0$, the mass mixing parameter $m_{12}^2 = 0$, and
$m_{22}^2 >0$, the second Higgs doublet $\phi_2$ does not acquire any
vev, and the SM scalar potential is recovered with the relation $v^2 =
v_1^2 = - m_{11}^2/\beta_1$.  This is the {\em inert doublet} scenario
with a perfectly $Z_2$ symmetric potential, in which all the
nonstandard scalars decouple from physical processes when the
parameter $m_{22}^2$ controlling their masses is taken to infinitely
large value.  Note that $m_{22}^2$, in this case, does not have its
origin in spontaneous symmetry breaking (SSB), and this is why its
large value could ensure decoupling. On the contrary, if we try to
establish the equivalence between the two parametrizations, given in
Eq.~(\ref{notation1}) and Eq.~(\ref{2hdm potential}), we have to go to
a situation when both the scalars receive vevs, and only then we
obtain the following relations:
\begin{eqnarray}
&& m_{11}^2 =-(\lambda_1v_1^2+\lambda_3v^2)~;~ m_{22}^2=
  -(\lambda_2v_2^2+\lambda_3v^2)~;~
  m_{12}^2=\frac{\lambda_5}{2}v_1v_2~;~
  \beta_1=2(\lambda_1+\lambda_3)~; \nonumber \\ &&
  \beta_2=2(\lambda_2+\lambda_3)~;~ \beta_3=2\lambda_3+\lambda_4~;~
  \beta_4=\frac{\lambda_5+\lambda_6}{2}-\lambda_4~;~
  \beta_5=\frac{\lambda_5-\lambda_6}{2}~.
\label{connections}
\end{eqnarray}
Note that when both the doublets receive vevs, one can trade the two
parameters $m_{11}^2$ and $m_{22}^2$ in favor of $v_1$ and $v_2$. Then
the magnitude of the third parameter $m_{12}^2$, or equivalently
$\lambda_5$, has nothing to do with SSB, and this parameter provides
the regulator whose large value ensures decoupling of all nonstandard
scalars from physical processes. However, while employing $m_{12}^2$
(or equivalently $\lambda_5$ in our parametrization) for decoupling,
one cannot escape from some tuning of parameters for softly broken
$Z_2$ as explained around Eq.~(\ref{kappaz2}), but no such tuning is
required for softly broken U(1) (discussed before).  Nondecoupling
would result when the symmetry of the potential is exact ($m_{12}^2 =
0$), {\em and at the same time}, both the scalars receive vevs (which
implies $\lambda_5 = 0$).  In this case all the non-SM physical scalar
masses would be proportional to the electroweak vev, and there is no
independent mass-dimensional parameter which has non-SSB origin.  As
illustrated in the inert doublet case, even with exactly symmetric
potential, decoupling is achieved in 2HDM.

Admittedly, the parametrization of Eq.~(\ref{2hdm potential}) is less
general than that of Eq.~(\ref{notation1}). Any connection between the
two sets of parameters can be established only when both the scalars
receive vevs.  The inert doublet scenario can be very easily realized
in the parametrization of Eq.~(\ref{notation1}), while just setting
$v_2 = 0$ in the parametrization of Eq.~(\ref{2hdm potential}) does
not lead us to the same limit. To appreciate this salient aspect, we
consider a simpler scenario when we have only one Higgs doublet. Then
the potential can be written in two equivalent ways: $V \sim \mu^2
|\phi|^2 + \lambda |\phi|^4$, and $V' \sim \lambda \left(|\phi|^2 -
v^2/2\right)^2$. They become truly equivalent when $\mu^2 < 0$, and
consequently, the scalar receives a vev. But when $\mu^2 > 0$, the
scalar remains inert. In that case, putting $v = 0$ in $V'$ does not
take us to the physical situation given by $V$, as the latter still
contains, in addition to $\lambda$, an independent dimensionful
parameter $\mu^2$. Our Eqs.~(\ref{2hdm potential}) and
(\ref{notation1}) are 2HDM generalizations of $V'$ and $V$,
respectively.

We note that Eq.~(\ref{2hdm potential}), where it is implicitly
assumed that both scalars receive vevs, covers a large class of 2HDM
models (Types I-IV), where a nonvanishing (and often large)
$\tan\beta$ has played an important r\^ole in addressing
phenomenological issues associated with processes like $b \to s
\gamma$, $B_s \to \ell^+ \ell^-$, $B \to D^{(*)} \ell \nu$,
etc~\cite{Branco:2011iw}.

To provide further intuition into the argument of decoupling and its
close connection to the existence of some non-SSB origin parameter, we
draw the following analogy.  It is well known that the top quark in
the SM does not decouple from $h\to\gamma\gamma$. This is because the
top quark receives all its mass from SSB and increasing the its mass
will invariably imply enhancing the Yukawa coupling ($h_t$). Now,
suppose that the top quark receives part of its mass ($M$) from some
non-SSB origin, i.e. $m_t=h_tv+M$.  Then the top-loop contribution
will yield a prefactor $h_tv/(h_tv+M)$. In this case, by taking $M \to
\infty$, the top quark contribution can be made to decouple from the
diphoton decay width of the Higgs boson.

\section{Three-Higgs-doublet models} 
$S_3$ or $A_4$ symmetric flavor models are typical examples which
employ three Higgs doublets.  With $\phi_1$, $\phi_2$ and $\phi_3$ as
the three scalar SU(2) doublets, the scalar potential for the $S_3$
symmetric case can be written as (see
e.g. \cite{Barradas-Guevara:2014yoa,Das:2014fea}, and also references
therein for flavor physics discussions both when the $S_3$ symmetry is
exact as well as when it is softly broken),
\begin{eqnarray}
V_{\rm 3HDM}^{S_3} & = & -\mu_1^2(\phi_1^\dagger\phi_1+\phi_2^\dagger\phi_2)-
\mu_3^2\phi_3^\dagger\phi_3 \nonumber \\
&& +\lambda_1 (\phi_1^\dagger\phi_1+\phi_2^\dagger\phi_2)^2 +\lambda_2
(\phi_1^\dagger\phi_2 -\phi_2^\dagger\phi_1)^2 +\lambda_3
\left\{(\phi_1^\dagger\phi_2+\phi_2^\dagger\phi_1)^2
+(\phi_1^\dagger\phi_1-\phi_2^\dagger\phi_2) ^2\right\} \nonumber \\
&& +\lambda_4
\left\{(\phi_3^\dagger\phi_1)(\phi_1^\dagger\phi_2+\phi_2^\dagger\phi_1)
+(\phi_3^\dagger\phi_2)(\phi_1^\dagger\phi_1-\phi_2^\dagger\phi_2) +
{\rm h. c.}\right\} \nonumber \\
&&
+\lambda_5(\phi_3^\dagger\phi_3)(\phi_1^\dagger\phi_1+\phi_2^\dagger\phi_2)
+ \lambda_6
\left\{(\phi_3^\dagger\phi_1)(\phi_1^\dagger\phi_3)+(\phi_3^\dagger\phi_2)
(\phi_2^\dagger\phi_3)\right\} 
\nonumber \\
&& +\lambda_7 \left\{(\phi_3^\dagger\phi_1)(\phi_3^\dagger\phi_1) +
(\phi_3^\dagger\phi_2)(\phi_3^\dagger\phi_2) +{\rm h. c.}\right\}
+\lambda_8(\phi_3^\dagger\phi_3)^2 \,.
\label{s3potential}
\end{eqnarray}
Assuming the lambdas to be real, potential minimization conditions
attribute a relation between two of the three vevs
($v_1=\sqrt{3}v_2$). Using this relation, an alignment limit can be
obtained for this model also \cite{Das:2014fea}.

Now we write the potential satisfying $A_4$ symmetry (see e.g. \cite{Ma:2001dn}),
\begin{eqnarray}
V_{\rm 3HDM}^{A_4} & = &
-\mu^2\left(\phi_1^\dagger\phi_1+\phi_2^\dagger\phi_2+\phi_3^\dagger\phi_3
\right)
+\lambda_1\left(\phi_1^\dagger\phi_1+\phi_2^\dagger\phi_2+\phi_3^\dagger\phi_3
\right)^2 \nonumber
\\ &&+\lambda_2\left(\phi_1^\dagger\phi_1\phi_2^\dagger\phi_2
+\phi_2^\dagger\phi_2\phi_3^\dagger\phi_3
+\phi_3^\dagger\phi_3\phi_1^\dagger\phi_1 \right)
+\lambda_3\left(\phi_1^\dagger\phi_2\phi_2^\dagger\phi_1
+\phi_2^\dagger\phi_3\phi_3^\dagger\phi_2
+\phi_3^\dagger\phi_1\phi_1^\dagger\phi_3 \right) \nonumber
\\ &&+\lambda_4\left[e^{i\epsilon}\left\{\left(\phi_1^\dagger\phi_2\right)^2+
\left(\phi_2^\dagger\phi_3\right)^2+\left(\phi_3^\dagger\phi_1\right)^2
\right\} + {\rm h.c.}\right]
\,.
\label{a4potential}
\end{eqnarray}
In one plausible scenario, the minimization conditions require that all the three
vevs are equal \cite{Toorop:2010ex}. This particular choice
automatically yields a SM-like Higgs as well as two pairs of complex
neutral states with mixed CP properties. Note that for $\epsilon=0$ in
\Eqn{a4potential}, the symmetry of the potential is enhanced to
$S_4$. However, our conclusions do not depend on the value of
$\epsilon$. 

Thus, a 3HDM can provide an SM-like Higgs along with two pairs of
charged scalars, as exemplified with $S_3$ and $A_4$ scenarios.  After
expressing the lambdas in terms of the physical masses, we obtain 
the following expressions for $\kappa_i~(i=1,2)$ in the alignment
limit, which are the same for both $S_3$ and $A_4$:
\begin{eqnarray}
\kappa_i = -\left(1+\frac{m_h^2}{2m_{i+}^2} \right)~~{\rm for}~i=1,2 \,.
\end{eqnarray}
Clearly, the charged scalars do not decouple from the diphoton decay
width, since $\kappa_i$ settles to $-1$ when $m_{i+}$ is very large
compared to $m_h$.  Note, both the charged scalars contribute in the
same direction to reduce $\mu_{\gamma\gamma}$.

Now we turn our attention to the case of a global continuous symmetry
in 3HDM potential. For illustration, we consider that the symmetry is
SO(2) under which $\phi_1$ and $\phi_2$ form a doublet.  The
expression for the scalar potential is similar to \Eqn{s3potential},
only that now $\lambda_4 =0$ and the potential contains an additional
bilinear term $(-\mu_{12}^2 \phi_1^\dagger\phi_2 + {\rm h.c.})$. The
real part of $\mu_{12}^2$ softly breaks the SO(2) symmetry and
prevents the occurrence of any massless scalar in the theory. In any
case, we assume $\mu_{12}^2$ to be real just like any other parameters
in the potential. The relevant minimization conditions are given by
\begin{subequations}
\begin{eqnarray}
v_1\mu_1^2+v_2\mu_{12}^2 &=& v_1(v_1^2+v_2^2)(\lambda_1+\lambda_3)
+\frac{1}{2}v_1v_3^2(\lambda_5+\lambda_6+2\lambda_7)\,, \\
v_2\mu_1^2+v_1\mu_{12}^2 &=& v_2(v_1^2+v_2^2)(\lambda_1+\lambda_3)
+\frac{1}{2}v_2v_3^2(\lambda_5+\lambda_6+2\lambda_7)\,.
\end{eqnarray}
\end{subequations}
Note that nonzero $\mu_{12}^2$ requires $v_1=v_2$. An interchange
symmetry ($1\leftrightarrow 2$) is accidentally preserved even after
spontaneous symmetry breaking.  We will have three CP even scalars
($h',~H,~h$), two pseudoscalars ($A_1,~A_2$) and two pairs of charged
scalars ($H_1^\pm,~H_2^\pm$). Among these, $h'$, $A_1$ and $H_1^\pm$
are odd under the interchange symmetry and the rest are even under
it. Being odd under this interchange symmetry, $h'$ does {\em not}
couple to gauge bosons as $h'VV$ ($V=W,~Z$). Appearance of such an
exotic scalar was noted earlier in the context of an $S_3$ symmetric
3HDM \cite{Bhattacharyya:2010hp,Bhattacharyya:2012ze,Das:2014fea}. The
soft breaking parameter ($\mu^2_{12}$) gets related to the mass of
$h'$ as
\begin{equation}
m_{h'}^2 = 2\mu_{12}^2 \,.
\end{equation}
It is straightforward to express the lambdas in terms of the physical
masses. We then obtain
\begin{subequations}
\begin{eqnarray}
\kappa_1 &=&
-\frac{1}{m_{1+}^2}\left(m_{1+}^2-m_{h'}^2+\frac{m_h^2}{2}\right)
\,, \label{kap1so2} \\ \kappa_2 &=& -\left(1+\frac{m_h^2}{2m_{2+}^2}
\right) \,.
\end{eqnarray}
\label{kapso2}
\end{subequations}
The similarity between \Eqn{kap1so2} and \Eqn{u1kappa} is
striking. Note that $(|m_{1+}^2 - m_{h'}^2|)$ is constrained from
unitarity.  Therefore, when the first charged Higgs mass $m_{1+}$ is
very large, $\kappa_1$ becomes vanishingly small.  However, this
decoupling does not occur in $\kappa_2$ which contains the second
charged Higgs mass $m_{2+}$.  It is not difficult to intuitively argue
that with an extended global symmetry SO(2)$\times$U(1), together with
an extra soft breaking parameter which is related to $m_{A2}$,
decoupling in $\kappa_2$ can be ensured.  Starting from the softly
broken SO(2) symmetric potential, this additional U(1) extension
($\phi_3 \to e^{i\alpha} \phi_3$) and its soft breaking can be
realized by putting $\lambda_7 = 0$ in Eq.~(\ref{s3potential}) and
introducing a term that softly breaks this U(1).  A crucial
observation we make in this paper is that the masses $m_A$ in the 2HDM
context and $m_{h'}$ in the 3HDM context enter into the expressions of
$\kappa_i$ -- e.g. see Eqs.~(\ref{u1kappa}) and (\ref{kap1so2}) --
only when they are related to {\em soft} global symmetry breaking
parameters.

\begin{table}[h]
\begin{center}
\begin{tabular}{|c|c|c|c|c|c|}
\hline
\multicolumn{2}{|c|}{Model} &  Expression for $\kappa_i$ & prediction $\mu_{\gamma\gamma}$ & prediction $\mu_{Z\gamma}$ & Decoupling? \\
\hline\hline
 & Softly broken $Z_2$ & $-\left(1+\frac{m_h^2}{2m_{1+}^2}-\frac{\lambda_5 v^2}{2m_{1+}^2} \right)$ & Depends on $\lambda_5$ & Depends on $\lambda_5$ & Possible \\
\cline{2-6}
2HDM & Exact $Z_2$ & $-\left(1+\frac{m_h^2}{2m_{1+}^2} \right)$ & $\le 0.9$ & $\le 0.96$ &No \\
\cline{2-6}
& Softly broken U(1) & $-\left(1+\frac{m_h^2}{2m_{1+}^2}-\frac{m_A^2}{m_{1+}^2} \right)$ & Depends on $m_A$ & Depends on $m_A$ & Yes \\
\hline
 & Exact $S_3$ & $-\left(1+\frac{m_h^2}{2m_{i+}^2} \right)$ for $i=1,2$ & $\le 0.8$ & $\le 0.93$ & No \\
\cline{2-6}
3HDM & Exact $A_4$ & $-\left(1+\frac{m_h^2}{2m_{i+}^2} \right)$ for $i=1,2$ & $\le 0.8$ & $\le 0.93$ & No \\
\cline{2-6}
& Softly broken SO(2) &
\specialcell{$\kappa_1=-\left(1+\frac{m_h^2}{2m_{1+}^2}-
\frac{m_{h'}^2}{m_{1+}^2} \right)$ \\
$\kappa_2=-\left(1+\frac{m_h^2}{2m_{2+}^2} \right)$}  & Depends on $m_{h'}$ & Depends on $m_{h'}$ & Partial \\
\hline
\end{tabular}
\end{center}
\caption{\em Behavior of 2HDM and 3HDM scenarios in the alignment
  limit strictly when all the doublets receive vevs. In the case of
  exact discrete symmetries, every charged scalar pair reduces
  $\mu_{\gamma\gamma}$ approximately by $0.1$. Although explicit
  expression for $\mu_{Z\gamma}$ is not shown in text, its predictions
  in different scenarios are displayed. In the last column where we
  say `Possible', we mean that decoupling can be achieved with some
  tuning, while in the last row `Partial' implies that only the first
  charged scalar decouples.}
\label{BSMtable}
\end{table}

\section{Conclusions and outlook} 
To our knowledge, this is the first attempt towards establishing a
connection between decoupling or nondecoupling of charged scalars from
the diphoton decay of the Higgs with the symmetries of the scalar
potential.  We show that charged scalars in multi-doublet scalar
extensions of the SM do not necessarily decouple from physical processes,
e.g. $\mu_{\gamma\gamma}$ in the context of this paper, specifically
when the potential has an exact symmetry and all the scalars receive
vevs. 

Here we give a few examples where phenomenology of such scenarios has
been studied. A spontaneously broken $Z_2$ symmetric potential in 2HDM
context, with a tiny but nonvanishing $v_2$, has been advocated to
account for the smallness of the neutrino mass and the stability of a
scalar dark matter on a cosmological scale \cite{Gabriel:2006ns}. A
few 3HDM examples are also in order.  Novel scalar sector
phenomenology with exotic scalar decay properties has been studied
with exact $S_3$ symmetric potential \cite{Bhattacharyya:2010hp,
  Bhattacharyya:2012ze}. General flavor physics studies were carried
out in $S_3$ \cite{Chen:2004rr} as well as in $A_4$
\cite{Toorop:2010ex} symmetric scenarios.

In such scenarios, a precisely measured $\mu_{\gamma\gamma}$ can smell
the presence of nonstandard scalars even if
they are super-heavy.  In fact, $\mu_{\gamma\gamma}$ can constrain the
{\em number} of such doublets. Table 1 shows that each additional pair
of charged scalars ($H_{i}^\pm$) reduces $\mu_{\gamma\gamma}$
approximately by 0.1 when the potential has an exact discrete
symmetry. Our illustrations are based on two- and three-Higgs-doublet
models which are motivated by flavor symmetries.  We have explicitly
demonstrated how soft breaking of a global U(1) symmetry can ensure
decoupling in 2HDM in the alignment limit. In the case of 3HDM, with a
softly broken global SO(2) symmetry in the potential, decoupling can
be ensured for one pair of charged scalars ($H_{1}^\pm$), while the
second pair ($H_{2}^\pm$) still do not decouple.  Employing the soft
breaking terms of an extended global continuous symmetry, namely,
SO(2) $\times$ U(1), the nondecoupling effects of $H_{2}^\pm$ can be
tamed. If we have more pairs of charged scalars in the theory stemming
from additional scalar doublets, even more enhanced or extended global
continuous symmetries $-$ only softly broken $-$ would be required to
ensure decoupling of all charged scalars from $\mu_{\gamma\gamma}$.
Keeping in mind the expected accuracy in the measurement of the $hhh$
vertex in the high luminosity option of LHC or in the future linear
collider, whose tree level expression in the alignment limit remains
the same as in SM even for multi-doublet Higgs structure,
$\mu_{\gamma\gamma}$ may offer a better bet for diagnosing the
underlying layers of the Higgs dynamics.

{\em To sum up}, if future measurement of $\mu_{\gamma\gamma}$ is
found to be consistent with the SM prediction to a high degree of
precision $-$ say better than 10\% $-$ non-inert type
multi-Higgs-doublet models with exact discrete symmetries will be
constrained.  We have demonstrated in this paper that the soft
breaking terms in the potential, which are often used in the
literature, can play an important r\^ole in ensuring decoupling, {\em
  albeit} with some tuning. To avoid this, one must start with a
global continuous symmetry in the potential followed by its soft
breaking.  In future, it would be interesting to explore the
consequences of global symmetries in the potential for nondoublet
scalar extensions in the present context.


\bibliographystyle{JHEP}
\bibliography{non_decoup.bib}
\end{document}